\documentclass[notitlepage,11pt,aps,tightenlines,unsortedaddress,nofootinbib,floatfix,showkeys]{revtex4-1}
\usepackage{newtxtext,newtxmath}
\usepackage{bm}
\usepackage{physics}
\usepackage[version=4]{mhchem}
\usepackage{mathtools,mathrsfs,amsfonts,dsfont}
\usepackage{relsize}
\usepackage{scalerel}
\usepackage{todonotes}
\usepackage{ulem}
\usepackage{xcolor}
\usepackage[unicode=true, pdfusetitle, bookmarks=true, bookmarksnumbered=false, bookmarksopen=false, breaklinks=true, pdfborder={0 0 0}, backref=false, colorlinks=true, linkcolor=blue, citecolor=blue, urlcolor=blue]{hyperref}
\usepackage{enumitem}
\usepackage{cancel}
\allowdisplaybreaks
\newcommand{\tpsi}{\skew{3}{\tilde}{\psi}}

\newcommand{\Ex}{\sm\vu{x}}
\newcommand{\Ey}{\sm\vu{y}}
\newcommand{\Ez}{\sm\vu{z}}
\newcommand{\sm}{\kern0.1em}
\newcommand{\smalldiv}{\raisebox{-0.2ex}{\resizebox{!}{1.6ex}{\sm/\sm}}}

\DeclareMathAlphabet{\mathbbold}{U}{bbold}{m}{n}

\begin{document}

\title{Detlef D\"urr, arrival-time distributions, and spin in Bohmian mechanics: Personal recollections and state-of-the-art}

\author{Siddhant Das}
\email{Siddhant.Das@physik.uni-muenchen.de}
\affiliation{Mathematisches Institut, Ludwig-Maximilians-Universit\"at M\"unchen, Theresienstr.\ 39, D-80333 M\"unchen, Germany}
\date{April.\ 30, 2022}

\begin{abstract}
I recount here my association with Prof.\ Detlef D\"urr leading to our memorable research collaboration on arrival-time distributions in quantum mechanics. He influenced my life, both personally and professionally, as few others have or ever will. Detlef is my role model for what a brilliant, discerning scientist, academic, and mentor can and should be.

The ``arrival-time problem" in quantum mechanics is examined selectively, with an emphasis on the arrival-time distributions of Bohmian particles. In what follows, the ``exotic'' Bohmian arrival-time distributions of spin-polarized electrons accelerating down a cylindrical waveguide [S.\ Das and D.\ D\"urr, \href{https://www.nature.com/articles/s41598-018-38261-40}{Sci.\ Rep.\ \textbf{9\sm}: 2242 (2019)}], and some variations thereof are discussed. I shall not go into the mathematical treatment more than is necessary to spell out the key results. The intention is to document the circumstances and motivations underlying the ideas.
\end{abstract}

\maketitle
\normalem

\begin{quote}
    \emph{``It is our choices, Harry, that show what we truly are, far more than our abilities.''} --- Albus Dumbledore (J.\ K.\ Rowling \cite{HP})
\end{quote}~\\[-15pt]

It was my good fortune to know Prof.\ Dr.\ Detlef D\"urr as a mentor, research collaborator, and close confidant from October 2016 until January 2021. We had regular interactions following his enriching and uniquely crafted course ``Introduction to Bohmian Mechanics,'' offered during the Summer Semester of 2016. This course attracted a large number of enthusiastic participants of all ages. As it happened, Detlef was officially teaching Bohmian mechanics for the very first time at LMU Munich.\footnote{And also the very last time, owing to his retirement in November of that year.} His lectures were generously seasoned with witty comments, historical anecdotes, and philosophical insight, while a palpable sense of seriousness was ever-present. By the end of the semester, his passion for Bohmian mechanics rubbed off on some of us. What more could one ask for?  

Bohmian mechanics (de Broglie-Bohm or pilot-wave theory, BM in the sequel) is an understanding of quantum physics featuring \emph{both} waves and particles, first formulated by Louis de Broglie in the 1920s. It was rediscovered, refined, and elucidated by David Bohm in the 50s, popularised and defended by John S.\ Bell in the 80s, and augmented by modern-day theorists. In BM, particles move in a highly non-Newtonian way, obeying the \emph{guiding equation}. This equation expresses their velocities in terms of the wave function, whose primary role in the theory is to guide particles along well-defined trajectories. Its statistical significance is a derived consequence, and it evolves by Schr\"odinger's equation at \emph{all} times.

This deterministic, matter-in-motion worldview of BM is at such odds with those suggested by orthodox/standard/textbook quantum mechanics (QM) that newcomers are typically and understandably perplexed. Detlef thus took great care in the BM course to set the stage. He began by reappraising material (handpicked from the ``red book'' \cite{DurrTeufel}) on dynamical systems, analytical mechanics, Boltzmann's statistical mechanics, and Brownian motion, emphasizing the conceptual precision of the ``classical'' theories over their known limitations in the quantum domain. The lectures on ``chance in physics'' \cite[Ch.\ 4]{DurrTeufel} were genuinely excellent and insightful.

BM proper was only discussed in the second half of the course. Except for the double-slit experiment, Detlef did not focus much attention on concrete examples/applications of the theory.\footnote{Nor did he address (or speculate on) the well-documented historical shortcomings responsible for physicists' disengagement from BM; see, e.g., \cite{becker,crossroads,Olival}.} He preferred to not lose sight of the forest (derivation of Born's rule, quantum nonlocality, measurement formalism) for the trees. If asked ``[i]s it in fact useful to compute Bohmian trajectories in various quantum mechanical situations we deem of interest?'', he candidly asserted, ``[i]n general, it is not!'' \cite[p.\ 155]{DurrTeufel}. Yet, Bohmian trajectories for benchmark QM problems, e.g., diffraction \cite{Wardhalfline}, two-slit interference (with \cite{Laloe} and without \cite{Dewdney} monitoring), barrier tunneling \cite{Norsentunneling,Dewdneytunneling}, EPR correlations \cite[Sec.\ 10.6]{BohmHiley}, etc., ``remain among the most striking illustrations so far of the insight provided by this theory into quantum phenomena'' \cite{Holland}. In particular, a principled treatment of the classical limit \cite{CL}, and a dynamical underpinning of the Pauli repulsion (or degeneracy pressure) \cite[p.\ 284]{HollandBook}, which these trajectories facilitate, impressed me early on.

It soon became evident, despite Detlef's tongue-in-cheek admonishment (``every lecture in quantum physics is an act of brainwashing''), that BM grounds the formalism and empirical predictions of standard QM coherently and compellingly; see \cite{BohmHiley,DurrTeufel,HollandBook,ShellyStanford,Bricmont,shellycushing,Oriols}. Fortunately, in so doing, ``it neither needs nor is embarrassed by an observer'' \cite[Ch.\ 15]{Bell}. Understanding how BM explains and computes the results of standard quantum experiments without deploying ill-defined measurement axioms \cite{AgainstMeasurement}, or a laundry list of quantum observables, was indeed a revelation.

One couldn't shake the feeling that ``the significance of the [quantum observables] has been exaggerated, in the sense that elements entering as useful \emph{mathematical techniques} have been raised to the level of fundamental concepts in the \emph{physical theory}'' \cite[p.\ 278]{BohmT}.\footnote{D.\ Bohm's close friend Richard P.\ Feynman couldn't have agreed more vehemently: ``If anyone tells me that `to every observable there corresponds a Hermitian operator for which the eigenvalues correspond to observed values,' \emph{I will defeat him! I will cut his feet off!}'' \cite{hestenes2003}.} Besides, ``[a]ny physicist who is ready to quantize everything under his pen'' \cite[p.\ vi]{DurrTeufel} must take the \emph{canonical quantization} procedure (for engineering quantum observables) with a grain of salt, as it is known (Groenewold-Van Hove theorem \cite{NOGO,NoGoRigorous}) that a procedure associating (i) every classical phase-space function \(f(x,p)\) to a quantum operator \(F\), in particular, (ii) \(x\) and \(p\) to the usual position and momentum operators of QM, and simultaneously (iii) classical Poisson brackets to quantum commutators: \(\{f,g\}\mapsto (1\smalldiv i\hbar)\sm[F,G]\), as Dirac envisioned, \emph{cannot} exist.\footnote{Ironically, Groenewold discovered his no-go result while attempting to demonstrate the impossibility of certain ``hidden variable'' accounts of quantum mechanics \cite{NOGO}, BM being the prime example of a ``hidden variable'' theory (see also \cite{RevModPhysBell}).} Indeed, in real-world applications, when the canonical quantization procedure leads to an impasse (see below), practitioners are forced to guess the relevant quantum observables from ad hoc symmetry principles, experience, or good taste. 

That being said, in all fairness, while operators as quantum observables play \emph{no} role in the formulation of BM, their effectiveness as empirical shorthands is best appreciated through a Bohmian lens. The \emph{emergence} of the ``momentum operator'' in a first principles Bohmian analysis of asymptotic momentum measurements \cite[Sec.\ 15.1.2.]{DurrTeufel} is a classic example (see, however, \cite{DGZOperators} for a general argument applicable also to generalized quantum observables as positive-operator-valued measures, or POVMs). More to the point, the concept of a particle trajectory is universally invoked in actual experiments when \emph{reconstructing} particle momentum from directly observable quantities, such as arrival (or flight) times. While the final momentum distribution agrees with the QM (also BM) prediction,\footnote{Although ``[i]n quantum mechanics there is no such concept as the path of a particle'' \cite[p.\ 2]{LandauLifshitz}.} the adequacy/limits of this time-honoured \emph{time-of-flight momentum spectroscopy} technique can only be analyzed within the context of a trajectory-containing quantum theory such as BM. 

Against this backdrop, and observing that the BM literature preemptively discusses experiments inspired by standard QM, I became interested in ``new'' experimental predictions of BM early on\footnote{As was, notably, Jean-Pierre Vigier \cite{JPV,JPV1,JPV2}, a prominent collaborator of L.\ de Broglie and D.\ Bohm.}---predictions that could help ``draw inferences from experiment regarding the particle track[s]'' \cite[Sec.\ 5.5]{HollandBook} within established limits \cite{DGZ}. But since there is a compelling argument \cite{DGZ} that BM is ``experimentally equivalent'' to regular QM, at least ``insofar as the latter is unambiguous'' \cite[Ch.\ 14]{Bell}, the prospects for definitive experiments were grim. Moreover, given that QM has been tested exhaustively, the existence of \emph{feasible} experiments (i.e., not of Wigner's-friend-type) for which its recipes would prove to be wanting and, concomitantly, the conceptual resources of BM could be brought to bear, was unprecedented. It must be mentioned that David Bohm speculated early on that the theory may ``lead to completely different kinds of predictions'' at distances ``of the order of \(10^{-13}\sm\)cm or less'' \cite{Bohm2}, though no concrete experimental ideas were outlined (see also \cite[Sec.\ 13]{BohmExp}). In addition, cosmological fingerprints of BM-type field theories suggested in \cite{AV} remain somewhat controversial.

\begin{figure}
    \centering
    \includegraphics[width=0.65\columnwidth]{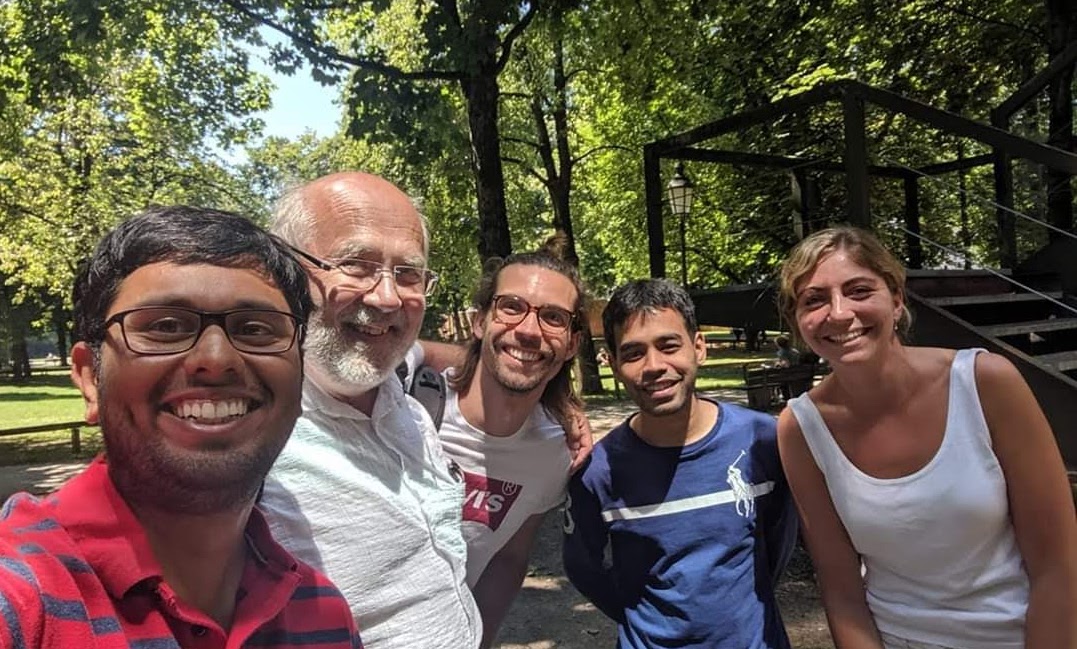}
    \caption{\footnotesize Detlef and his disciples at the Mensa Arcisstra{\ss}e (circa July 2019). From left, Arun Ravishankar, Detlef D\"urr, Nikolai Leopold, Siddhant Das, and Manuela Feistl.} 
    \label{fig1}
\end{figure}

A silver lining appeared in the final week of the BM course when Nikolai Leopold (see Fig.\ \ref{fig1}) briefly introduced the ``arrival-time problem'' in QM, which I was completely ignorant of. I feel certain that my worldline would have been very different had I missed Nikolai's talk. He was a Ph.D.\ student in Detlef's ``Workgroup Mathematical Foundations of Physics'' at the time, working on ``Effective Evolution Equations from Quantum Mechanics'' \cite{NickThesis}. Nikolai was exploring as a side project the parameter space of an arrival-time experiment involving photons. This experiment, requiring a coherent superposition of two Gaussian wave packets, was previously considered by Nicola Vona in his Ph.D.\ thesis \cite{NicolaPhd}.\footnote{Recently, Rafsanjani \emph{et al.}\ \cite{AliJavad} have suggested a very practical realization of this type of experiment using a double-slit interferometer.} At the same time, an experimental implementation of this work was being discussed with the group of Prof.\ Harald Weinfurter, whose expertise in quantum optics meant the calculations had to be specifically tailored for photons.

Arrival-time or time-of-flight (ToF) experiments are scattering experiments concerned with ``\emph{when} the detector clicks'' rather than ``\emph{where} the detector clicks.'' The latter is a mainstay of standard QM (and quantum field theory) with specialized theoretical tools at its service, e.g., the \(S\)-matrix formalism. The former, on the other hand, has been the subject of much controversy since nearly the beginning of QM.\footnote{In \cite[Nov.\ 17, 1925]{Pauliquote}, Wolfgang Pauli wrote in a letter to Niels Bohr: ``Perhaps the following points in the right direction of possible further progress. In the new theory [matrix mechanics], not all physically observable quantities really occur, namely, the time instants of the transition processes are still missing, which are certainly observable in principle (e.g., the instants of ejections of electrons in the photoelectric effect).''} In a typical theoretical analysis of a ToF experiment, one considers a particle with a well-localized wave function \(\psi_0(\vb{r})\) at time zero,\footnote{For example, the moment of releasing a particle from a trap after state preparation.} propagating either freely or in specified external potentials. Critical to any such discussion is the probability \(\smash{\Pi(\tau)\,d\tau}\) that the particle's time-of-arrival on a \emph{specified surface} \(\mathcal{D}\) (representing the surface of a detector, e.g., a scintillation screen) is between times \(\smash{\tau}\) and \(\smash{\tau+d\tau}\).\footnote{Not to be confused with the probability of locating the particle in a volume element \(d^3r\) surrounding \(\vb{r}\) at a \emph{specified time} \(t\), viz., \(|\psi_t(\vb{r})|^2\sm d^3r\) (Born's rule!), where \(\psi_t\) is the particle's wave function at time \(t\).} Today, the computation of the probability density \(\Pi(\tau)\), which is empirically well-accessible, is one of the last areas where physicists disagree about what QM should predict \cite{MUGA1,MSP,backwall}. This, in a nutshell, is the ``arrival-time problem''.

ToF experiments had long been on Detlef's radar, considering how straightforward it was to describe them using the inventory of BM. (In particular, the particle trajectories defined by the guiding equation.) That BM is naturally suited for handling arrival- and tunneling-time problems has long been recognized \cite{DDGZ,Leav,Leav98}, \cite[Sec.\ 5.5]{HollandBook}. But Detlef's interest in this subject went back to his work on quantum scattering formalism \cite[Ch.\ 16]{DurrTeufel}, \cite{DGZScattering,Vona2015,DTManyParticle,DurrMB}, especially the flux-across-surfaces theorems \cite{DDGZ,DDGZ96,FAS} and the time-dependent analysis of Gamow decay \cite{Gamow,*GamowEr,GrummtPhd,Grummt2021}. A grounding in scattering theory enabled him to appreciate the significance of the probability current (or quantum flux) density in ToF experiments early on. Before I met Detlef, relevant research on arrival times was conducted in collaboration with his former students Helmut Satzger \cite{Helmut}, Nicola Vona \cite{NicolaPhd}, and G\"unter Hinrichs \cite{Vona1}. So, in retrospect, it is somewhat surprising that he did not bring up arrival-time experiments earlier in the course when pressed about scenarios where BM would be particularly useful.

The issue of arrival times got under my skin. For an outstanding problem in QM and an area of considerable theoretical activity (see below), the conspicuous lack of experimental contact or guidance seemed especially striking. That situation seems to have changed very little even today (see, however, \cite{DoubleSlit}).\footnote{In contrast, the ``tunneling-time problem,'' which asks ``how long does it take for a particle to tunnel?'' has received much experimental attention in recent years \cite{Attoclock,TT,Steinberg,Steinberg1,GField}.} One possible explanation is that the available ToF predictions have not previously been worked out in sufficient detail to be compared with actual experiments.

ToF measurements, on the other hand, are central to experimental methods determining energies and momenta in both single- and few-particle scattering experiments featuring wave packet coherence \cite{fitting}, including, e.g., the temperature of trapped ions \cite{FerdiThermal}, band structure \cite{bandstructure,bandstructure3}, Wigner's function \cite{Pfau}, chemical reaction dynamics (as in the Rydberg tagging ToF technique \cite{RydbergTOF2}), etc. However, Newtonian or geometric optics \cite{GO} \emph{Ans\"atze} frequently suffice for interpreting the ToF measurements in these instances.\footnote{This is hardly surprising from a Bohmian standpoint as the appearance of Newtonian behaviour under scattering conditions is a common characteristic of BM \cite{SarahRomer}. As a result, in many cases, the arrival times of Bohmian particles may be approximated via Newtonian assumptions.} The ``unreasonable effectiveness'' of semiclassical heuristics in laboratory ToF experiments has, however, also played a key role in \emph{masking} the underlying concerns posed by the ``arrival-time problem''. 

I had a strong urge to improve matters on this front. As it happened, given Detlef's approaching retirement, his limited time shared between existing Ph.D.\ students, and some health-care needs of his family, he had somewhat reluctantly decided not to supervise me for a master's thesis. In the meantime, I continued corresponding with Nikolai about his work and tried suggesting ways to relax various idealizations in the theoretical treatments. I soon realized that the association with Nikolai was a bit like ``killing two birds with one stone.'' First, I saw Detlef more often, long after the BM course had ended, and second, I could focus on the arrival-time problem that had so enticed me. As time went on, Detlef's attention and interest in the problem seemed to be reinvigorated. I soon became an integral part of the discussions and even started developing the ideas in ways I felt might be more promising---something both Detlef and Nikolai welcomed very graciously.

During Detlef's retirement celebrations in November, Nikolai introduced me to Prof.\ Roderich Tumulka (or Rodi, as he is fondly called) over lunch, who had already independently worked on the quantum ``arrival-time problem'' \cite{ABC,ABC1,ABCMulti}. This meeting turned out to be a critical turning point. After briefly listening to what I was pursuing, Rodi suggested that if I was going to make the effort to analyze realistic ToF experiments, it should be refocused on massive particles rather than photons (Bohmian theories for photons were somewhat contentious\footnote{For this reason I believe the impact of the weakly measured average photon trajectories \cite{Kocsis} was slightly diminished, even though ``trajectories measured in this fashion reproduce those predicted in the Bohm-de Broglie interpretation of quantum mechanics.'' A similar experiment with Argon atoms was ongoing a few years ago \cite{BarkerWM}.}). This proved to be invaluable advice, for which I remain grateful to him. That evening, I stopped working on photon arrival times and started hunting for feasible experiments with massive particles. 

Before continuing my personal story further, let me make a few observations about certain theoretical efforts surrounding the ``arrival-time problem''. Various ToF distributions suggested in the literature can be divided into two broad categories. We limit our discussion here to the category of ideal (or intrinsic) arrival-time distributions that are \emph{apparatus-independent} theoretical predictions given by some functional of the initial wave function \(\psi_0\) and the geometrical surface of the detector \(\mathcal{D}\) (typically a single point on a line in one-dimensional discussions). The second category of non-ideal (or measurement-inspired) ToF distributions typically features simplified detector models characterized by new phenomenological parameters. Various suggestions have been put forward, e.g., simple absorbing boundary conditions \cite{Werner,ABC}, complex potentials \cite{histories1,currentB,Allcock2}, wave function collapse (both detector-induced \cite{zeno, JN} and spontaneous \cite{EEQT1,EEQT2}), path integrals with absorbing boundaries \cite{PI1,PI2}, a variety of quantum clocks \cite[Ch.\ 8]{MUGA}, and even a timeless formulation of quantum measurement \cite{Maccone,Pullin}.

In the category of ideal ToF distributions---insofar as such a description may be warranted---those based on arrival-time {\emph {observables}} are widely studied in the literature. The observables in question are typically motivated via classical arrival-time expressions, e.g., the well-known Aharonov-Bohm operator \cite{AhBohm,HPaul}
\begin{equation}\label{tauAB}
    T_{\text{AB}} = \frac{m}{2}\sm\Big[(L-X)P^{-1}+P^{-1}(L-X)\Big],
\end{equation}
obtained by a symmetric quantization of the classical ToF expression \(\smash{\tau_{\text{cl}}(x,p) = m\sm (L-x)\smalldiv p}\) of a freely moving particle that at \(\smash{t=0}\) had position \(x\) and momentum \(p\), and arrived after time \(\tau_{\text{cl}}\) at a distant point \(L\) on a line.\footnote{Though it is rarely recognized that the correct classical ToF formula valid for an arbitrary initial point \(\smash{(x,p)}\) in phase space is \(\smash{\tau_{\text{cl}}(x,p) =m\sm (L-x)\smalldiv p}\) for \(\smash{\text{sgn}\sm p=\text{sgn}\sm(L-x)}\) and \(\infty\) otherwise; the quantization of which seems nothing short of impossible.} In \eqref{tauAB}, \(X\) and \(P\) are the standard position and momentum operators of QM, respectively.

A preliminary observation that justified interest in this kind of operator was that \(T_{\text{AB}}\) is (formally) canonically conjugate to the free-motion Hamiltonian, raising expectations for the long-sought ``time-energy uncertainty relation''. However, \(T_{\text{AB}}\) is not a self-adjoint operator \cite[p.\ 60]{PauliThm}, hence fails to provide a quantum observable \`a la Dirac and von Neumann. As it happens, this is not a showstopper because the (doubly degenerate) eigenstates of this operator, denoted \(\ket{\tau,\pm}\), can be used to define a generalized observable or POVM \cite{MugaBadBad,MugaBad}, and in turn the arrival-time distribution
\begin{equation}\label{PAB}
    \Pi_{\text{AB}}(\tau) = |\!\braket{\tau,+}{\psi_0}\!|^2+|\!\braket{\tau,-}{\psi_0}\!|^2,\qquad \braket{\tau,\pm}{\psi_0}=\int_0^{\infty}\!\!\!dp\,\sqrt{\frac{p}{2\sm\pi\hbar\, m}}\,\tpsi_\tau(\pm\sm p)\,e^{\pm\sm i pL/\hbar},
\end{equation}
where \(\smash{\tpsi_t(p)=\braket{p}{\psi_t}}\) is the momentum-space representation of the freely-evolved wave function \(\psi_t(x)\). Several authors arrived at this ToF distribution (see \cite{gaugeinv} for a critical review), notably J.\ Kijowski \cite{kijToF}, whose \emph{axiomatic} derivation also applies to a three-dimensional setting for an infinite, planar \(\mathcal{D}\). 

The above method of generating arrival-time operators from classical expressions seldom works for scenarios other than free, one-dimensional motion, owing to the fact that analogues of \(\tau_{\text{cl}}(x,p)\) are rarely available for motion in external potentials. This has resulted in a rather long, circuitous, and near endless proliferation of (inequivalent) arrival-time observables \cite{HM,HMM,Leon,Baute,Baute1,Galaponantithesis,Galaponsad,Goto1,Goto2,STS1,BF,causality,circleop}, not to mention the many technical and interpretative issues accompanying them\footnote{For example, the suggestion of assigning probabilities to ``negative arrival times'' (\(T_{\text{AB}}\) has negative eigenvalues), which causes the predicted mean ToF, \(\matrixel{\psi_0}{T_{\text{AB}}}{\psi_0}\), to vanish for \emph{any} real-valued \(\psi_0\) regardless of its form, or of the distance from the detector \cite{gaugeinv}.} discussed at length in \cite{gaugeinv,Mielnik}. To this end, a gauge-invariant generalization of \(T_{\text{AB}}\) applicable in the presence of external magnetic fields is yet to be found, see \cite[Sec.~3]{gaugeinv}. In light of these issues the value of a theory such as BM, whose framework offers a starting point completely different from relying on guessing a self-adjoint operator or POVM, becomes evident.

We turn now to the quantum flux distribution
\begin{equation}\label{qf}
    \Pi_{\text{QF}}(\tau) =
     \int_{\mathcal{D}}\vb{j}(\vb{r},\tau)\cdot d\vb{s},
\end{equation}
where \(\vb{j}(\vb{r},t)\) is the quantum flux (or probability current) density associated with Schr\"odinger's equation, and \(\smash{d\vb{s}}\) the surface measure at \(\smash{\vb{r}\in\mathcal{D}}\). As indicated earlier, this distribution would be a natural guess for an arrival-time distribution from scattering considerations (see also \cite[p.\ 6]{gaugeinv}), and prima facie has a very wide domain of applicability including three-dimensional settings, generic detection surfaces \(\mathcal{D}\) and, most importantly, ToF experiments featuring external electromagnetic fields. However, it ceases to be a meaningful ToF distribution for some wave functions, particularly those for which \eqref{qf} is negative.\footnote{For this reason, the quantum flux distribution does not follow from a POVM \cite{Vona1}.} This problem is remedied by the Bohmian arrival-time distribution, to which we turn next.

Numerous authors have discussed the arrival-time distribution of Bohmian particles \cite{DDGZ,Grubl}, \cite[Sec.\ 5.5]{HollandBook}, most notably C.\ Richard Leavens \cite{Leav,Leav98}, \cite[Ch.\ 5]{MUGA}. To specify this ToF distribution, consider a particle with position \(\vb{R}\), obeying the de Broglie-Bohm (guiding) equation 
\begin{equation}\label{guide} 
 \dot{\vb{R}}(t)=\vb{v}_{\text{dBB}}^{\psi_t}\big(\vb{R}(t),t\big),
\end{equation}
where \(\dot{\vb{R}}\) denotes the particle's velocity and \(\psi_t\) its wave function, a solution of Schr\"odinger's equation:
\begin{equation}\label{sch}
    i\hbar\,\frac{\partial \psi_t}{\partial t} = H\psi_t.
\end{equation}
The precise form of the velocity field \(\vb{v}_{\text{dBB}}^{\psi_t}(\vb{r},t)\) and Hamiltonian \(H\) depends on particle spin, external potentials, and on whether the motion is relativistic or non-relativistic \cite{BohmHiley,DurrTeufel,HollandBook,ShellyStanford,Bricmont,shellycushing,Oriols}. As soon as these are specified, particle trajectories may be calculated by solving the dynamical equations subject to suitable initial conditions \(\vb{R}_0\) and \(\psi_0\).

With these theoretical resources and only these, the \emph{first} arrival time (or hitting time) of a Bohmian trajectory \(\vb{R}(t)\), starting at some \(\smash{\vb{R}_0 \in {\text{supp}}(\psi_0)}\)\footnote{The support of \(\psi_0\), {\text{supp}}(\(\psi_0\)), is the region for which \(\psi_0\) is nonzero.} at time zero, and arriving at \(\mathcal{D}\) at time \(\tau\), can be defined as follows:
\begin{equation}\label{tau}
    \tau(\vb{R}_0) = \text{inf}\sm\big\{\, t\ge 0\sm:\sm \vb{R}(t)\in\mathcal{D}\,\text{ and }\,\vb{R}(0)=\vb{R}_0\in\text{supp}(\psi_0)\sm\big\}
\end{equation}
---a quantity that has no counterpart in standard QM.\footnote{Or, for that matter in either spontaneous collapse (wave-function-reduction) or many-world theories.} In standard QM, the detection event is tied to the observation-induced collapse of the wave function, whose time of occurrence (relevant to the measured ToF) is ambiguous. N.B: The infimum in \eqref{tau} ensures that the ToF of Bohmian particles not intercepting \(\mathcal{D}\), e.g., those back-scattered at an intervening potential barrier giving rise to non-detection events, is infinity, since \(\text{inf}\sm\emptyset:=\infty\).

Now, given some statistical hypothesis about the initial conditions \(\vb{R}_0\) and \(\psi_0\), the Bohmian arrival-time distribution \(\Pi_{\text{BM}}(\tau)\) can be determined using \eqref{tau}. In a sequence of identically prepared ToF experiments with a fixed initial wave function \(\psi_0\), the \(\vb{R}_0\)s are typically \(|\psi_0|^2\)-distributed,\footnote{In such an ensemble, the Bohmian particle positions remain $|\psi_t|^2$-distributed at \emph{any} later time $t$ by virtue of the guiding equation. This property, dubbed \emph{equivariance}, expresses the compatibility between the evolution of the wave function given by Eq.\ \eqref{sch} and the evolution of the particle's position given by Eq.\ \eqref{guide}.} therefore, the probability distribution of \eqref{tau} can be written as
\begin{equation}\label{PiBM}
    \Pi_{\text{BM}}(\tau) = \int_{\text{supp}(\psi_0)}\!\!d^3\kern-0.1em R_0~\,\delta\big(\tau-\tau(\vb{R}_0)\big)\sm|\psi_0(\vb{R}_0)|^2,
\end{equation}
where \(\delta(\cdot)\) is Dirac's delta function. For completeness, a {\emph {non-detection probability}}
\begin{equation}\label{PinfBM}
    P_{\text{BM}}(\infty) = \int_{\tau^{-1}(\infty)}\!\!\!d^3\kern-0.1em R_0~|\psi_0(\vb{R}_0)|^2
\end{equation}
is defined, which is the \(|\psi_0|^2\)-measure of precisely those initial conditions for which the Bohmian particle does {\emph {not}} intercept \(\smash{\mathcal{D}}\) at any finite time. It follows that, together with definition \eqref{PinfBM}, the integral of \eqref{PiBM} for all \(\smash{\tau\ge0}\) evaluates to one, as it must for any well-behaved probability distribution. (Equations (\ref{PiBM}-\ref{PinfBM}) can be easily adapted to instances where \(\psi_0\) is chosen at random from a statistical mixture encoded in a density matrix.)

The distribution \(\Pi_{\text{BM}}\) is well-defined for generic external electromagnetic (even time-dependent) potentials as well as generic (finite, non-finite, even multipiece) detection surfaces \(\mathcal{D}\). It also naturally generalizes to multi-particle arrival-time problems unlike \(\Pi_{\text{QF}}\), see \cite{DDSE,DurrMB,DTManyParticle}, as well as to a joint probability distribution of impact positions \emph{and} arrival times \cite{DoubleSlit}. These attributes explain why the Bohmian trajectories are widely used for addressing arrival- and tunneling-time problems in the literature; see \cite{DoubleSlit,DDSE,Wuhan,Leavens1996,Leavensspeedup,Nicolas,Oriolstunneling,Mousavi,Mousavitunnel,BohmbeatsKij,Home,CPviol,HomeCP,Golshani,Demir,GField,Nogami} for many applications over the years.

A word of caution: the Bohmian first-arrival times discussed here, which do not take into account the physical interaction with the detectors (characteristic of other ideal arrival-time distributions discussed above), concern first and foremost the arrival times of Bohmian \emph{trajectories}, rather than the times registered by macroscopic measurement devices. Whether the former are good approximations to the latter demands further study. In most familiar settings the agreement is highly persuasive; see, e.g., \cite{DGZScattering,DoubleSlit}. However, It is conceivable that the differences between the two may be significant in some cases. In such circumstances, a deeper Bohmian analysis that goes well beyond the pragmatic assumptions underpinning \(\Pi_{\text{BM}}\) would be required.

When I began working with Detlef and Nikolai in 2017, I was unfamiliar with much of the literature on ToF distributions, but \(\Pi_{\text{BM}}\) ``smelled right'' and was an easy starting point for studying ToF experiments utilizing BM. Yet, from a practical standpoint, it is rarely directly computable\footnote{However, in one-dimensional settings at least, an explicit formula \cite[Eq.\ (12)]{Kreidl} for \(\Pi_{\text{BM}}\) has been directly computed.} except, of course, when it reduces \emph{exactly} to the quantum flux distribution \(\Pi_{\text{QF}}(\tau)\) under the current positivity condition (CPC) \cite[p.\ 6]{gaugeinv}:
\begin{equation}\label{CPC}
    \forall\sm t>0\quad\text{and}\quad\forall\sm\vb{r}\,\in\,\mathcal{D},\quad \vb{j}(\vb{r},t)\cdot d\vb{s}\sm\ge\sm0. 
\end{equation}
Note that whenever \eqref{CPC} holds, \(\Pi_{\text{QF}}(\tau)\) is guaranteed to be a meaningful probability distribution. If the CPC does \emph{not} hold, a closed-form expression for \(\Pi_{\text{BM}}\) involving only the wave function \(\smash{\psi_t}\) and/or its partial derivatives cannot, in general, be given; one must therefore evaluate it numerically. However, it turns out that \eqref{CPC} is practically always met in the far-field or scattering regime and is central to the flux-across-surfaces theorems mentioned earlier \cite{DDGZ,DDGZ96,FAS}.

Since \(\Pi_{\text{QF}}\) \emph{is} the Bohmian arrival-time distribution in the absence of backflow, Detlef sometimes preferred to express \(\Pi_{\text{BM}}\) itself as a flux integral over \(\mathcal{D}\), \`a la Eq.\ \eqref{qf}, introducing a so-called \emph{truncated flux} density \(\tilde{\vb{j}}\) \cite[Eq.\ (10)]{DDGZ}, implicitly defined in terms of the first arrivals of Bohmian trajectories at \(\mathcal{D}\). Despite being mathematically identical to \eqref{PiBM}, the truncated-flux formulation of \(\Pi_{\text{BM}}\) is a bit too formal\footnote{I do not mean to imply here that Detlef was a formalist. For the most part he ``stayed well clear of purely academic questions like, for instance, the domain of self-adjointness of observable operators'', firmly believing that ``whoever sees in those a key to understanding quantum mechanics is certainly wrong'' \cite[p.\ 217]{DurrDustin}.} for most physics students, and is not especially suited for numerical calculations; I therefore typically avoid using it.

This ``cornucopia of existing ToF predictions,'' of which we have introduced but a few, was seen by Detlef as a clear sign that the ``chapter on time measurements in quantum mechanics'' is still substantially unwritten. ``Quantum mechanics, arguably the most thoroughly tested theory of nature, surely deserves better,'' he once declared emphatically. Evidently, in order to draft this unwritten chapter, we must \emph{do} ToF experiments using present-day technology and not just theorize about them. Detlef, of course, was well aware that discerning competing ToF predictions in actual experiments presents several technical and practical challenges.

Foremost among them was one stemming from the early observation that ``several theoretical treatments converge to essentially the same arrival-time distribution at asymptotic distances and times from the particle source and intervening scatterers'' \cite[p.\ 358]{MUGA1}. So, in practical experiments probing the far-field regime, the Bohmian arrival-time predictions for massive particles would be extremely difficult to distinguish from competing predictions, particularly those based on semiclassical (or, better, mixed quantum-classical) considerations, such as \cite{WardNico}
\begin{equation}\label{Psc}
      \Pi_{\text{SC}}(\tau) = \left(\frac{m}{\tau}\right)^3\!\int_{\mathcal{D}}\,\left|\tpsi_0\!\left(\frac{m\kern0.1em\vb{r}^\prime}{\tau}\right)\right|^2\frac{\vb{r}^\prime\cdot d\vb{s}}{\tau},
\end{equation}
implicit in the ToF momentum measurements mentioned above.\footnote{Before I became involved, Detlef and Nicola had been actively seeking suitable experimental scenarios for distinguishing \(\Pi_{\text{SC}}\) from the quantum flux (equivalently, Bohmian) ToF distribution \cite[Ch.\ 4]{NicolaPhd}.}

Considering this, one is generally advised to relocate ``the detectors closer to the region of coherent wave packet production, or closer to the interaction region'' \cite[p.\ 419]{MUGA1}, i.e., to devise \emph{near-field} arrival-time measurements, where semiclassical approximations are invalid and discernible differences between different theoretical proposals are manifest. When actual experimental parameters are considered, it rapidly becomes clear that the particle should arrive nearly instantly at the detector; therefore, resolving ToF distributions at short distances demands impeccable time-resolution capabilities.

\emph{Back to the story.} Drawing on this background, Detlef proposed that it might be fruitful to investigate (the considerably more challenging) few-particle arrival-time problem, where the quantum flux distribution was inapplicable \cite{DTManyParticle} but a natural many-body generalization of \(\Pi_{\text{BM}}(\tau)\) would still be usable.\footnote{I should note in passing that there is presently much interest in experimentally probing the interplay of Coulomb repulsion and the degeneracy pressure in few-electron experiments \cite{Batelaan,Hommelhoff}---a problem Detlef was particularly interested in.}

I subsequently began examining a two-particle arrival-time problem. Most ideas had to be fleshed out from scratch because I could find no earlier work on few-body arrival-time problems based on BM. At times I managed to overwork myself, attempting to coax the theory into revealing something unexpected, to the point that Detlef often intervened with remarks like ``Siddhant, don't forget to live'' \ldots I must add that Detlef enthusiastically supported my applications for \emph{DAAD-Abschlussstipendium}, the \emph{LMU-Scholarship}, as well as the \emph{Bayern State Scholarship}, all of which offered valuable financial support during the period 2016-2017. It cannot be overstated how wildly efficient Detlef was at drafting grant applications, writing recommendation letters, referee reports, rebuttals, etc., and concurrently taking great care of his many pet animals (no pun intended).\footnote{Prof.\ Tim Maudlin has shared an amusing anecdote about how Detlef once politely excused himself in the middle of someone's talk, reasoning that it was time to walk his dog.}

A few weeks went by with no tangible progress. Enter a second, watershed moment (the first being the already stated counsel from Rodi): an opportunity to give a talk at the group ``Oberseminar.'' Detlef encouraged me to present a brief update on my current work at this weekly gathering, which Detlef and his associates took very seriously at the time. I was quite terrified by the invitation since I had really nothing to report; no results, nothing! Detlef convinced me that doing an overview presentation and providing some perspective on the literature would suffice. But, at that time I was too impatient to closely engage with the vast ``arrival-time problem'' literature, and I was also afraid of committing the scandal of delivering a dull seminar.

\begin{figure}
    \centering
    \includegraphics[width=\textwidth]{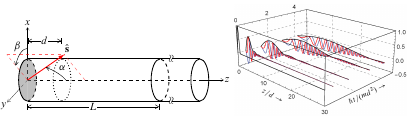}
    \caption{\footnotesize\emph{Left panel:} A semi-infinite cylindrical waveguide mounted on the \(xy\)-plane of a right-handed coordinate system, the axis of the cylinder defining the \(z\)-axis. A spin-1/2 particle is prepared in a ground state of a cylindrical box of width \(d\). This state is spin-polarized along the unit vector \(\vu{s}\). At time zero, the particle is released into the waveguide leading to detection events on a detection plate (at \(\smash{z=L}\)) downstream, where its arrival times are recorded. \emph{Right panel:} Snapshots of the function \(W(z,t)\): \(\mathrm{Re}[W]\) (red), \(\mathrm{Im}[W]\) (blue), \(|W|\) (black) for select values of~\(t\).}
    \label{fig2}
\end{figure}

Therefore, one afternoon I decided to undertake something that seemed far more interesting, and likely manageable in about a week's time. The seminar was to be held on July 12, 2017. I focused my attention on outlining a single-particle ToF experiment that could be presented to the group members in under an hour. The precise experimental setting I considered is as follows: A spin-1/2 particle is confined to move along a long waveguide, modelled as a semi-infinite cylinder. It is first trapped between the end face of the waveguide and an impenetrable potential barrier placed at a distance $d$, as shown in Fig.\ \ref{fig2}. The particle is prepared in the ground state \(\psi_0\) of this cylindrical box, then the barrier at \(d\) is suddenly switched off at \(\smash{t=0}\), enabling the particle to move freely within the waveguide. The arrival surface \(\mathcal{D}\) is the plane situated a distance \(\smash{L~(\gg d)}\) from the end face of the waveguide, where arrival times are recorded. This experiment undergoes several repetitions under identical conditions, yielding the arrival-time distribution \(\Pi(\tau)\).

In what follows, I describe the main motivations that underpin this particular experimental setup. First, as discussed in \cite{Lamb}, trapping the particle in a box and then abruptly releasing it was intended to produce a well-defined and dependably preparable wave function,\footnote{If the barrier at \(d\) is released sufficiently rapidly, the particle's wave function immediately after is \(\psi_0\).} as well as a well-defined starting time for the ToF experiment. Since only the front wall is removed rather than both walls, the wave packet expands unidirectionally rather than bidirectionally. In the latter instance, two detector plates would be required, otherwise, detection events in one of the directions would go unreported, contributing to an appreciable non-detection probability. 

The waveguide was incorporated in order to allow the particle to travel a long distance, thereby facilitating \emph{far-field} detections, while concomitantly preventing scattering-regime characteristics from setting in. The latter manifest whenever the potentials are of short range. This, I hoped, would provide measurable differences between \(\Pi_{\text{BM/QF}}\) and, say, \(\Pi_{\text{SC}}\), given that these distributions normally agree when the Bohmian trajectories become asymptotically Newtonian. 

Finally, why specify spin-1/2 particles? Introducing spin into the discussion may appear to some readers an unnecessary complication, as spin is easily ignored in most situations. The main reason for considering particles with nonzero spin here was that the Bohmian guiding equation, Eq.\ \eqref{guide}, for a spin-1/2 particle, is more-or-less uniquely pinned down by relativistic covariance considerations \cite{Holland92,Holland,Hollanduniqueness}, whereas there was still some concern about the uniqueness of the guiding equation for spin-0 particles \cite{Deotto}.\footnote{These issues arise because the goal of BM is to underpin the statistical predictions of QM with deterministic dynamics of point particles, and there are a priori far too many options for formulating such theories.}

The Bohmian velocity field (cf.\ Eq.\ \eqref{guide}) for spin-1/2 particles of mass \(m\) moving at non-relativistic speeds, adapted here to the special case of space-spin factorized or spin-polarized wave functions (those of the form \(\Psi_t = \psi_t(\vb{r}) \chi\) with ``spatial part'' \(\psi_t\) and ``spin part'' \(\chi\)---a two-spinor), is given by \cite[Ch.\ 10]{BohmHiley}
\begin{equation}\label{guide1}
    \vb{v}_{\text{dBB}}^{\psi_t}(\vb{r},t) = \frac{\hbar}{m}\sm\Im\kern-0.1em\left[\frac{\pmb{\nabla}\psi_t}{\psi_t }\right]+\frac{\hbar}{m}\sm\Re\kern-0.1em\left[\frac{\pmb{\nabla}\psi_t }{\psi_t }\right]\times\vu{s},
\end{equation}
where \(\vu{s}:=\chi^\dagger\bm{\sigma}\chi\) is a unit vector defined with the help of Pauli's spin-vector matrix \(\bm{\sigma}=\sigma_x\Ex+\sigma_y\Ey+\sigma_z\Ez\). This velocity formula differs from that for a spin-0 particle by an additional term, the so-called ``spin term'', which is unfortunately often omitted in most presentations of spin in BM, perhaps following an early proposal of John\ S.\ Bell \cite[pp.\ 171-181]{BellSpinTerm}.\footnote{This term, it turns out, is entirely responsible for electron mobility in the ground state of a hydrogen atom \cite{Colijn,Colijn2,Colijn1}, and it may even lead to particle trajectories crossing the symmetry axis in a double-slit experiment \cite{Holland}. Finally, the so-called ``ring currents'' arising out of this term are central to the study of aromaticity, see \cite{aromaticity}.}

Let me spell out a few steps involved in a Bohmian treatment of the proposed experiment featuring a spin-dependent guiding equation. For concreteness, I chose a harmonic potential \(V_\perp(x,y)=m\sm \omega^2\sm(x^2+y^2)\smalldiv2\) to model the waveguide, which acts on the wave function at all times. The barriers at \(\smash{z=0}\) and \(\smash{z=d}\) were modelled by the (time-dependent) potential \(\smash{V_\parallel(z)=v(z)+\theta(-\,t)\sm v(d-z)}\), where \(\smash{v(z)=0}\) for \(\smash{z>0}\) and infinite otherwise, and \(\theta\) is Heaviside's unit step function that eliminates the barrier \(v(d-z)\) for \(\smash{t>0}\).

In units for which \(\smash{\hbar=m=d=1}\),\footnote{In other words, when masses, lengths, and times are expressed in units of \(m\), \(d\), and \(\smash{md^2\smalldiv\hbar}\), respectively.} the parameter space of this problem is comprised of the trapping frequency \(\omega\), the flight-path-length \(L\), and two ``spin-polarization angles'', \(\alpha\), \(\beta\), specifying the orientation of \(\vu{s}\) (see Fig.\ \ref{fig2}).

The ground state wave functions of the trapped spin-1/2 particle assume a space-spin factorized form \(\Psi_0(\vb{r})=\psi_0(\vb{r})\chi\), with spatial and spin parts given by 
\begin{equation}\label{initspinor}
    \psi_0(\vb{r})=\sqrt{\frac{2\sm \omega}{\pi}}\,\theta(z)\theta(1-z)\sin(\pi z)\,e^{-\frac{\omega}{2}(x^2+y^2)},\quad\text{ respectively }\quad\chi=\mqty(\cos\frac{\alpha}{2}\\[5pt]e^{i\beta}\sin\frac{\alpha}{2}).
\end{equation}
These wave functions vanish in the regions \(z\le0\) and \(z\ge d\) (\(\smash{=1}\) in our units), differing only by the values of the spin-part parameters \(\smash{\alpha\in[0,\pi]}\), \(\smash{\beta\in[0,2\sm\pi)}\).\footnote{This particular parameterization of \(\chi\) is done in order that \(\vu{s}:=\chi^\dagger\bm{\sigma}\chi=\sin\alpha\big(\cos\beta\Ex+\sin\beta\Ez\big)+\cos\alpha\Ez\).} Lacking time, I chose the ground states corresponding to \(\smash{\alpha=0}\), \(\smash{\alpha=\pi}\) and \(\smash{\alpha=\pi\smalldiv2}\), with \(\smash{\beta=0}\), for the seminar. With these, we obtain what may be referred to as the ``spin-parallel'', ``spin-antiparallel'' and ``spin-perpendicular'' wave functions, respectively, which determines the orientation of the unit (spin) vector \(\vu{s}\) relative to the waveguide axis associated with these wave functions. 

Once the barrier is turned off, a prepared ground state wave function spreads dispersively into the waveguide, obeying Eq.\ \eqref{sch} with Hamiltonian \(\smash{H=(1\smalldiv2)\sm\laplacian+V_\perp(x,y)+v(z)}\). It continues to remain space-spin factorized with ``spin part'' \(\chi\), the same as that set at time zero, and ``spatial part'' also identical to \(\psi_0\), Eq.\ \eqref{initspinor}, except that \(\theta(z)\theta(1-z)\sin(\pi z)\) is replaced by \(\theta(z)W(z,t)\sm e^{-\sm i\omega t}\). I could obtain this function \(W(z,t)\) in closed-form \cite[Eq.\ (11)]{DD} by solving Schr\"odinger's equation, which led to a considerable simplification in the numerical treatment of the arrival-time experiment. A few snapshots of \(W(z,t)\) are depicted in Fig.\ \ref{fig2} (right panel). It shows the unfolding of a truly remarkable wave phenomenon called \emph{diffraction in time} \cite{Moshinsky}, manifesting in response to a sudden change in the boundary conditions of the wave function at a given surface (in this case, the plane \(\smash{z=1}\)). Observe that the prepared wave function breaks into an infinite collection of tiny ripples or wavelets close to this surface, nucleating into \emph{self-similar} wave packets, each propagating down the waveguide in succession.

This wave function determines the particle velocity in the guidance equation \eqref{guide1}, in turn, the Bohmian trajectories \(\smash{\vb{R}(t)=X(t)\sm\Ex+Y(t)\sm\Ey+Z(t)\sm\Ez}\), which can be used to express the first arrival-time, \eqref{tau}, of an individual trajectory, \emph{viz.}, \(\smash{\tau(\vb{R}_0) = \text{inf}\sm\big\{\, t\ge 0\sm:\sm Z(t)=L\,\text{ and }\,\smash{0<Z_0<1}\sm\big\}}\), as a function of the initial position \(\vb{R}_0\) and the flight-distance \(L\). And, assuming that the initial positions are randomly distributed according to the \(|\psi_0|^2\)-distribution,\footnote{Note that \(\smash{\Psi_0^\dagger\Psi_0^{\phantom{\dagger}}\overset{\eqref{initspinor}}{=}|\psi_0|^2}\) for any \(\alpha\), \(\beta\).} the desired Bohmian ToF distribution is obtained via Eq.\ \eqref{PiBM}. Since the parameters in the initial wave function are, in view of \eqref{initspinor}, \(\alpha\) and \(\beta\),  we denote \(\Pi_{\text{BM}}(\tau)\) by \(\Pi^{\alpha|\beta}_{\text{BM}}(\tau)\). 
 
\begin{figure}
    \centering
    \includegraphics[width=\textwidth]{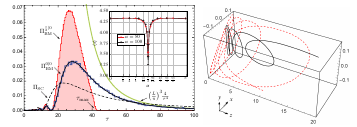}
    \caption{\footnotesize\emph{Left panel}: Bohmian arrival-time distributions for particles spin-polarized parallel \(\Pi^{0|0}(\tau)\) and perpendicular \(\Pi^{\frac{\pi}{2}|0}(\tau)\) to the waveguide axis. The solid, dashed curves are the quantum flux (QF) resp.\ semiclassical (SC) ToF predictions. Arrival times on the horizontal axis are reported in the units of \(md^2/\hbar\). Each histogram has been generated with $10^5$ Bohmian trajectories. We see agreement between $\Pi^{0|0}(\tau)$ and $\Pi_{\text{QF}}(\tau)$. The distribution $\Pi^{\frac{\pi}{2}|0}(\tau)$ predicts no first arrivals after time $\tau_{\text{max}}$. Note the disagreement of all distributions with $\Pi_{\text{SC}}(\tau)$. \emph{Left panel Inset:} Mean first-arrival-time \(\expval{\tau}\) vs.\ spin orientation angle \(\alpha\) for \(\smash{L=10}\) and \(\smash{\beta=0}\). \emph{Right panel}: Typical Bohmian trajectories for \(\smash{\alpha=\pi/2}\), \(\smash{\beta=0}\) (solid) and \(\smash{\beta=3\pi/4}\) (dashed). Both trajectories are plotted for \(\smash{t\in[0,5]}\) with (common) initial condition \(\vb{R}_0=0.05\sm(\vu{x}+\sqrt{3}\sm\vu{y})+0.5\sm\vu{z}\). Here, \(\smash{z=1}\) denotes the hard-wall potential barrier, which was switched off at \(\smash{t=0}\).}
    \label{fig3}
\end{figure}

Histograms of the BM distributions for spin-parallel \(\smash{(\alpha=\beta=0)}\) and spin-perpendicular \((\alpha=\pi/2,\sm\beta=0)\) to the waveguide-axis are shown in Fig.\ \ref{fig3} for the parameters \(\smash{\omega=10^3}\) and \(\smash{L=100}\). To obtain these graphs, I sampled \(\approx 10^5\) random initial positions from the \(|\psi_0|^2\)-distribution, solved the Bohmian guiding equations numerically\footnote{The assistance of Grzesio Gradziuk and Leopold Kellers with the numerical simulations using Mathematica was essential and greatly appreciated, and provided me with an opportunity to learn Mathematica programming for numerical calculations.} for each point in this ensemble, continued until the trajectory hit \(\smash{z=L}\), recorded the arrival time and plotted the histogram for \(\Pi_{\text{BM}}(\tau)\). 

For the spin-parallel (also antiparallel) wave function, \(\Pi^{0|0}_{\text{BM}}(\tau)\) (blue histogram in Fig.\ \ref{fig3}) agrees with \(\Pi_{\text{QF}}\) (i.e., Eq.\ \eqref{qf} evaluated with the Pauli-current \cite{Mike}) owing to the CPC \eqref{CPC} being satisfied. (This was an important consistency check that the numerical calculations were on track.) It could be shown that this ToF distribution is independent of the trapping frequency \(\omega\), which sets the effective diameter of the waveguide \(\smash{\approx \sqrt{\hbar\smalldiv m\omega}}\), and has an asymptotic \(\tau^{-4}\) fall-off.\footnote{A direct consequence of the back-wall at \(\smash{z=0}\), without which it would have an asymptotic \(\tau^{-2}\) fall-off.}

On the other hand, for the spin-perpendicular wave function, the results of the simulations were striking!\ In this case, no Bohmian particle struck \(\smash{z=L}\) beyond a characteristic maximum arrival-time \(\tau_{\text{max}}\) regardless of the initial position \(\vb{R}_0\). Consequently, the density \(\Pi^{\frac{\pi}{2}|0}_{\text{BM}}(\tau)\) vanished for \(\smash{\tau>\tau_{\text{max}}}\), strongly disagreeing with the QF distribution. This was because, to my complete surprise, \emph{the CPC failed to apply for this wave function}. (Typical Bohmian trajectories for this wave function are depicted in Fig.\ \ref{fig3}, right panel.) Furthermore, this feature remained intact for all values of \(L\) and \(\omega\), except when \(\smash{\omega=0}\), in which case there is no radial confinement. 

In fact, in the limit \(\smash{\omega\to0}\), \(\smash{\Pi_{\text{BM}}^{\alpha|\beta}=\Pi_{\text{QF}}}\) for \emph{all} \(\alpha,\sm \beta\) hence, in particular, \(\smash{\tau_{\text{max}}\to\infty}\), and the spin-polarization-dependence is lost. This demonstrates the importance of the waveguide confinement to observing backflow (i.e., the violation of CPC), and in turn the spin-dependence of \(\Pi_{\text{BM}}\).\footnote{While the waveguide does play a crucial role in provoking and even influencing the amount of backflow in this setup, it would manifest neither for a spin-0 particle moving within the waveguide nor for a spin-1/2 particle moving freely.} (For a discussion of the limit \(\smash{\omega\to0}\), see \cite[Sec.\ V. C.]{Exotic}.)

Finally, a mention should be made of the smaller secondary lobes of the ToF distributions (visible for \(\tau\lesssim 20\) in Fig.\ \ref{fig3}). They are created by a particle propagating within the support of the smaller wavelets created at the moment the barrier at \(d\) is removed, as explained previously and seen in Fig.\ \ref{fig2} (right panel). Because of the nodes separating these wave packets, the Bohmian particle remains inside the support of the wave packet for which its random initial position was realized at \(\smash{t=0}\). The wave packets advance in time carrying the particle with them, hence the arrival times are recorded in bunches. Such secondary lobes caused by diffraction in time have been detected experimentally in the ToF distributions of atoms \cite{DIT} and neutrons \cite{DIT1}.

So, serendipitously, I had stumbled upon something completely unexpected. The results would not have been expected based on the wave function alone, since the Hamiltonian describing the particle's propagation through the waveguide is \emph{spin-independent}.\footnote{Although in quantum chemistry one often encounters spin-dependent answers to questions for which the Hamiltonian could well be spin-independent \cite[p.\ 38]{Walters}.} The spin-dependent arrival-time distributions found in this setup are a genuine consequence of the particle's dynamics determined by Eqs.\ \eqref{guide} and \eqref{guide1}.\footnote{We note in passing that spin-dependent arrival-time distributions would also be predicted by some of the absorbing Boundary Conditions (ABCs) \cite{ABCDirac}, but there it would not be particularly unexpected since a spin-dependent boundary condition is explicitly imposed on the wave function at \(\smash{z=L}\).}

Needless to say, participants at the seminar were unpreparedly surprised. Of course, Detlef did not buy into the perplexing results right away, but he was maximally pleased by the disagreement with the semiclassical prediction (dashed curve in Fig.\ \ref{fig3}) which drops asymptotically as \(\tau^{-2}\).\footnote{Actually, \eqref{Psc} is only applicable for free propagation, so it had to be generalized in order to account for the presence of the waveguide as well as the back wall; see \cite[Eq.\ (18)]{DD}.} Soon after this oberseminar both Detlef and I worked very intensely on this arrival-time setup, slowly learning a great deal about the role of spin in the process. We had regular weekly meetings (every Wednesday). He was very attentive to details and always had a listening ear. 

Subsequently, a short article detailing these findings was published \cite{DD}, most of which transformed into my master's thesis that I defended on October 25, 2017. (Detlef personally invited Prof.\ Weinfurter, who graciously probed me until 19:00 that evening, despite the fact that the defence began at 16:15.) In terms of conveying research, Detlef's philosophy was ``one idea per paper.'' He was also a very talented writer\footnote{With a healthy appetite for Latin idioms, e.g., ``cum grano salis'', ``experimentum crucis'', etc.}, capable of masterfully intertwining physics with formalism, as anybody who has read his many books and research articles can attest.

Such ``unexpected and very well-articulated arrival-time distributions,'' Detlef insisted, ``almost demand experimental inspection.'' Therefore, in the meantime, we also contacted a number of experimental groups to realize this arrival-time experiment.

As I was wrapping up my master's program, Detlef invited me to continue working on arrival times as his doctoral student and arranged funds to support me. This is when I earned the distinction of being his last Ph.D.\ student. Following a short break with my family in India, I returned to Munich in mid-2018 and resumed our collaboration. Over the following two years, a very detailed understanding of the waveguide setup and variants thereof was acquired, thanks especially to the contributions of Markus N\"oth---a gifted mathematical physicist, close friend, and collaborator.

The predicted spin-dependent arrival-time distributions would readily manifest for various kinds of confining potentials in waveguides, and in waveguides of different cross-sections, e.g., elliptical ones. In addition, it was found that a parallel plate waveguide suppresses the arrival-time distributions for a significantly greater range of \(\alpha\)s about \(\pi\smalldiv2\) than a circular one, making it the most promising arrangement for experimentally probing the spin-dependent distributions.\footnote{I worked out the parallel-plate setup to answer Ward Struyve's concern that the suppression in arrival times in the cylindrical waveguide setup manifests extremely close to \(\smash{\alpha=\pi/2}\), as evidenced by the sharp drop in the \(\expval{\tau}\) vs.\ \(\alpha\) curve shown in Fig.\ \ref{fig3} (left panel inset).} These results also do not depend on the presence or precise form of the back wall. They hold for particles of spin \(\smash{\ge1/2}\) and, somewhat surprisingly, for composite particles such as individual hydrogen atoms prepared in triplet states as well. Similar results arise in a relativistic treatment based on Dirac's equation and the Bohm-Dirac guiding equation, whose validity extends to relativistic Bohmian velocities \cite{DasDirac}. Interestingly, nodes in the initial wave function \(\psi_0\) induce \emph{gaps} (or ``no arrival windows'') in the distribution for \(\Pi^{\frac{\pi}{2}|0}_{\text{BM}}(\tau)\). A striking illustration of this occurs if one prepares the particle in excited states \(\smash{\propto\sin(n\pi)}\) instead of the ground state \eqref{initspinor}. It will take some time to prepare all of these extensions for publication, and work is currently underway.

In the middle of this euphoric collaboration, all hell was unleashed by the Covid-19 pandemic.\ Detlef's untimely passing on January 3, 2021, was a seismic event for me. I recall with some sadness a moment in December 2019 when he said to me as we were waiting for a train to Frankfurt that, in his opinion ``I had put my finger on something very profound,'' and that he ``would not be around when they see it.'' In retrospect, these were truly crushing words, and I had no clue that the unthinkable would come to pass so soon. I had developed such a close bond with him, and so many aspects of my life had been shared with him, that it was of course very difficult to move on. To whom would I turn in pursuing research now? It seemed for some time afterwards to be a pointless endeavour.

Our work had come to a point where he had begun one of his presentations by saying ``I shall report on a recent result by Siddhant Das for arrival time distributions in QM, which may provide the elements for a paradigm change to a quantum world which is understandable'' \cite{MLL}. I eventually came to realize that for me to give up at this stage would seem a betrayal of both the confidence he had in me, and in the work we had done together. I suppose that the primary reason I continue it is that, though I loved Detlef immensely, I love the pursuit of physics research with the same passion that he had for it, even more. I am very thankful for the support of my family and friends, many of them former students and colleagues of his (I would mention especially Anirudh Chandrasekaran, James M.\ Wilkes, Serj Aristarhov, Markus N\"oth, Paula Reichert, Ward Struyve, Dirk Deckert, Dustin Lazarovici, Prof.\ Hemalatha Thiagarajan, Prof.\ Subhendra D.\ Mahanti, Prof.\ Jean Bricmont, Prof.\ A.\ Shadi Tahvildar-Zadeh, Prof.\ Sheldon Goldstein, and Prof.\ Tim and Vishnya Maudlin), who have helped me hold my personal and academic life together and slowly resume work.

Clearly, much remains to be done to turn these preliminary findings into more realistic experimental proposals, and there may well be more surprises to come in the future (some that Detlef may already have anticipated). Only time will tell, of course. Reflecting on this time in my life, I consider myself extremely privileged to have been a small part of Detlef's scientific legacy. His warmth, contagious enthusiasm, collegial grace, and, above all, his willingness to welcome nascent scientists to the field have improved our community in ways that will not be forgotten.
\newpage
I thank James M.\ Wilkes and Dustin Lazarovici for carefully reviewing this article and for their invaluable editorial input. Thank you also to Stephen N.\ Lyle for pointing out a few typos.

\bibliography{DurrRef}
\end{document}